\documentclass[twocolumn]{aastex631}

\newcommand{\etal}{\textit{et al.~}}
\newcommand{\eV}{\mathrm{eV}}
\newcommand{\EeV}{\mathrm{EeV}}
\newcommand{\Mpc}{\mathrm{Mpc}}
\newcommand{\km}{\mathrm{km}}
\providecommand{\degree}{^\circ}

\usepackage{amsmath}
\usepackage{mathtools}
\usepackage{lipsum}

\shorttitle{Indications of a Cosmic Ray Source in the Perseus-Pisces Supercluster}
\shortauthors{The Telescope Array Collaboration}
\graphicspath{{./}{figures/}{myFigures/}}
\begin{document}

\title{Indications of a Cosmic Ray Source in the Perseus-Pisces Supercluster}

\correspondingauthor{Jihyun Kim}
\email{jihyun@cosmic.utah.edu}

\collaboration{200}{The Telescope Array Collaboration}

\author[0000-0001-6141-4205]{R.U. Abbasi}
\affiliation{Department of Physics, Loyola University Chicago, Chicago, Illinois, USA}

\author[0000-0001-5206-4223]{T. Abu-Zayyad}
\affiliation{Department of Physics, Loyola University Chicago, Chicago, Illinois, USA}
\affiliation{High Energy Astrophysics Institute and Department of Physics and Astronomy, University of Utah, Salt Lake City, Utah, USA}

\author{M. Allen}
\affiliation{High Energy Astrophysics Institute and Department of Physics and Astronomy, University of Utah, Salt Lake City, Utah, USA}

\author{Y. Arai}
\affiliation{Graduate School of Science, Osaka City University, Osaka, Osaka, Japan}

\author{R. Arimura}
\affiliation{Graduate School of Science, Osaka City University, Osaka, Osaka, Japan}

\author{E. Barcikowski}
\affiliation{High Energy Astrophysics Institute and Department of Physics and Astronomy, University of Utah, Salt Lake City, Utah, USA}

\author{J.W. Belz}
\affiliation{High Energy Astrophysics Institute and Department of Physics and Astronomy, University of Utah, Salt Lake City, Utah, USA}

\author{D.R. Bergman}
\affiliation{High Energy Astrophysics Institute and Department of Physics and Astronomy, University of Utah, Salt Lake City, Utah, USA}

\author{S.A. Blake}
\affiliation{High Energy Astrophysics Institute and Department of Physics and Astronomy, University of Utah, Salt Lake City, Utah, USA}

\author{I. Buckland}
\affiliation{High Energy Astrophysics Institute and Department of Physics and Astronomy, University of Utah, Salt Lake City, Utah, USA}

% ...



\begin{abstract}
The Telescope Array Collaboration has observed an excess of events with 
$E \ge 10^{19.4} ~\eV$ in the data which is centered at (RA, dec) = ($19\degree$, $35\degree$). 
This is near the center of the Perseus-Pisces supercluster (PPSC). The PPSC is 
about $70 ~\Mpc$ distant and is the closest supercluster in the Northern Hemisphere 
(other than the Virgo supercluster of which we are a part). A Li-Ma oversampling 
analysis with $20\degree$-radius circles indicates an excess in the arrival 
direction of events with a local significance of about 4 standard deviations. 
The probability of having such excess close to the PPSC by chance is estimated 
to be 3.5 standard deviations. This result indicates that a cosmic ray source 
likely exists in that supercluster.
\end{abstract}



\section{Introduction} \label{sec:intro}
Ultrahigh energy cosmic rays (UHECR) are energetic particles originating 
from outer space, having energies greater than $10^{18} ~\eV$, that impinge on the 
Earth's atmosphere. Their sources remain one of the most important questions 
to be answered. The main areas of UHECR study are their energy spectrum, 
mass composition, and searches for anisotropy, which are expected to shed 
light on the question of sources. The Telescope Array (TA) hotspot \citep{TA+hotspot:2014} 
which is about $20\degree$ in radius and centered at ($146.7\degree, 43.2\degree$) 
is one indication of anisotropy. Other previous signs are the Auger dipole 
\citep{Auger+dipole:2017}, most of the evidence for which is in the southern 
hemisphere, and the Auger excess \citep{Auger+ani:2015} near the location of 
Centaurus A, the closest active galactic nucleus. The nearby Virgo cluster, 
dominated by M87, does not appear as an excess in UHECR data.

The TA collaboration hotspot, reported in 2014 \citep{TA+hotspot:2014}, was 
found in the arrival direction of UHECR events with energy greater than $57 ~\EeV$ 
in the northern sky. In that publication, an oversampling analysis using the 
intermediate angular scale of $20\degree$-radius circles was conducted for the 
first 5 years of data collected with the TA scintillator surface detector array. 
The maximum excess appeared at the position ($146.7\degree, 43.2\degree$) in 
equatorial coordinates with a Li-Ma significance of $5.1\sigma$. The probability 
of having such an excess by chance was estimated via Monte Carlo (MC) studies 
to be $3.4\sigma$. However, there are no known prominent sources aligned with 
the hotspot within the cosmic ray events' horizon. There have been correlation 
studies between the hotspot and possible sources \citep{He:2016,Kim:2019}; however, 
the source of the hotspot remains inconclusive. The hotspot persists in 
TA data with over $3\sigma$ significance \citep{TA+ICRC:2021}.

In the Northern Hemisphere, the next closest component of the local large-scale 
structure is the Perseus-Pisces supercluster (PPSC). It stretches across the sky 
from the Perseus cluster to the Pegasus cluster and contains about 16 galactic 
clusters and groups containing tens of thousands of galaxies.

In this paper, we report a new excess of events at slightly lower energies than 
the original hotspot. While studying the spectrum mismatch above $10^{19.5} ~\eV$
seen in the TA and Auger data \citep{TA+Declination:2021}, we made sky maps of events 
for energy ranges, $E \geq 10^{19.4} ~\eV$, $E \geq 10^{19.5} ~\eV$, and $E \geq 10^{19.6} ~\eV$. 
The maximum Li-Ma significance appeared at the location of the PPSC for each energy 
range. We now present these three data sets.

The structure of this paper is as follows. In Section \ref{sec:data} we describe 
the TA experiment and the data used in this study. Section \ref{sec:analysis}
has three parts. First, we present the Li-Ma oversampling analysis using $20\degree$-radius 
circles to determine the significance of the excess of events in arrival directions. 
Then we show sky maps made with the representative elements of the PPSC. Finally, 
we describe the MC simulation used to estimate the chance probability of having 
an excess close by the prominent local large-scale structures. In Section \ref{sec:summary} 
we summarize our findings.

\section{TA experiment and the data} \label{sec:data}
The TA is the largest observatory for UHECRs in the Northern Hemisphere. The 
observatory is approximately 1400 m above sea level, and it is centered at 
$39.3\degree$N and $112.9\degree$W, in the west desert of Utah, USA. It consists 
of a surface detector (SD) array and three fluorescence detector (FD) stations 
viewing the sky over the array. It is designed for observing extensive air showers 
induced by UHECRs using a hybrid technique.

The SD array consists of 507 plastic scintillation detectors deployed on a square 
grid with $1.2 ~\km$ spacing, covering a total area of $\sim$700 $~\km^2$ \citep{TASD:2012}. 
They measure the shower footprint and lateral distribution at the Earth’s surface. 
Three FD stations, instrumented with 38 telescopes, are situated at the apices of 
a triangle with each station having a field of view overlooking the area of the 
SD array \citep{TAFD:2012}. The FDs are suitable for measuring the longitudinal 
shower development of events to estimate the mass of the primary particle; however, 
their data are limited to $\sim$10\% duty cycle since they operate only on moonless 
nights. In this study, we focus on the data recorded by the SD array to take full 
advantage of its high duty cycle (greater than 95\%). 

For this work, we used the data collected between May 11 of 2008 and May 10 of 
2019—a data-taking period of 11 years. The event selection criteria that apply 
to the data are as follows:\\
\indent 1. Energy $\geq 10^{19.4}$ eV,\\
\indent 2. Zenith angle of arrival direction $< 55\degree$,\\
\indent 3. At least five SDs triggered,\\
\indent 4. Shower geometry and lateral distribution function fit $\chi^2/{\rm dof} < 4$,\\
\indent 5. Reconstructed pointing direction error $< 5\degree$,\\
\indent 6. The fractional uncertainty of the energy estimator $S_{800} < 25$\%,\\
\indent 7. The largest signal counter is surrounded by four working counters: 
there must be at least one working counter to the left, right, down, up on the 
grid of counter with the largest signal. These counters do not have to be 
immediate neighbors of the largest signal counter.

These are our standard selection criteria that have been used previously for 
anisotropy studies. The energy of reconstructed events is determined by the SD 
$S_{800}$ which is then renormalized by 1/1.27 as previously determined to match 
the SD scale to the calorimetrically determined FD energy scale \citep{TA+Espectrum:2013}. 
A total of 864 events meet these selection criteria. The energy and angular 
resolution of events are from 10\% to 20\% and $1.0\degree$ to $1.5\degree$, 
respectively \citep{TA+coldspot:2018}.

\section{Li-Ma oversampling analysis} \label{sec:analysis}
In conducting the spectrum study where this work originated \citep{TA+Declination:2021}, 
part of our investigation was whether the hotspot extends down into lower declinations 
at slightly lower energies. Therefore, we adopted $20\degree$ angular windows for oversampling 
to be consistent with the original hotspot analysis method \citep{TA+hotspot:2014}. As a result, 
we found new excesses of events in the distribution of arrival directions.

The TA observes from $90\degree$ to $-15.7\degree$ in declination 
and $0\degree$ to $360\degree$ in right ascension. To estimate the %isotropic 
background event rate, $10^5$ events were generated within that field of view, 
assuming an isotropic flux and taking into account the geometrical exposure of 
the SD. The statistical significance of the excess of the data was compared to 
the isotropic events at each grid point, using $0.1\degree \times 0.1\degree$ 
grid spacing. The significance was calculated utilizing the Li-Ma method \citep{LiMa:1983}:

\begin{figure}[h]
\begin{equation}
    \centering
    \scalebox{1}{$S_{\rm LM}=\sqrt{2}\,\left[ N_{\rm on} \ln{ \left( \frac{(1+\alpha) N_{\rm on}} 
    {\alpha(N_{\rm on}+N_{\rm off})} \right) }+N_{\rm off} \ln{ \left( \frac{(1+\alpha) N_{\rm off}}{N_{\rm on}+N_{\rm off}} \right)} \right]^{1/2}$},
    \label{equ}       
\end{equation}
\end{figure}

\noindent where $\alpha = N_{\rm sim,circle}/(N_{\rm sim,total} - N_{\rm sim,circle})$. 
Here, $N_{\rm sim,total}$ is the total number of isotropic background events we generated, 
$N_{\rm sim,total} = 10^5$, $N_{\rm sim,circle}$ means the number of isotropic events inside 
the $20\degree$-radius circle at each grid point, $N_{\rm on}$ represents the number of 
observed events inside the circle at the same grid point, and 
$N_{\rm off} = N_{\rm data,total} - N_{\rm on}$, where $N_{\rm data,total}$ is the total 
number of data in each data set. The number of background events inside the $20\degree$-radius 
circle is estimated by the exposure ratio $\alpha$ and the data, 
$N_{\rm bg} = \alpha \, N_{\rm off}$.

When compared to the isotropic background events, new Li-Ma significance results are 
calculated for each energy cut and are summarized as follows. For $E \geq 10^{19.4} ~\eV$, 
the maximum Li-Ma significance appears at ($17.4\degree, 36.0\degree$) with $4.4\sigma$. 
There are 85 out of 864 observed events within the central $20\degree$-radius circle 
oversampling, whereas 49.5 events are expected as background from the isotropic assumption. For 
$E \geq 10^{19.5} ~\eV$, the maximum Li-Ma significance appears at ($19.0\degree, 35.1\degree$) 
with $4.2\sigma$. There are 59 out of 558 observed events within the central $20\degree$-radius 
circle oversampling, whereas 31.5 events are expected as background from the isotropic assumption. 
For $E \geq 10^{19.6} ~\eV$, the maximum Li-Ma significance appears at ($19.7\degree, 34.6\degree$) 
with $4.0\sigma$. There are 39 out of 335 observed events within the central 
$20\degree$-radius circle oversampling, whereas 18.6 events are expected as background from the 
isotropic assumption.

\begin{figure*}[t]
    \centering
    \includegraphics[width=\textwidth]{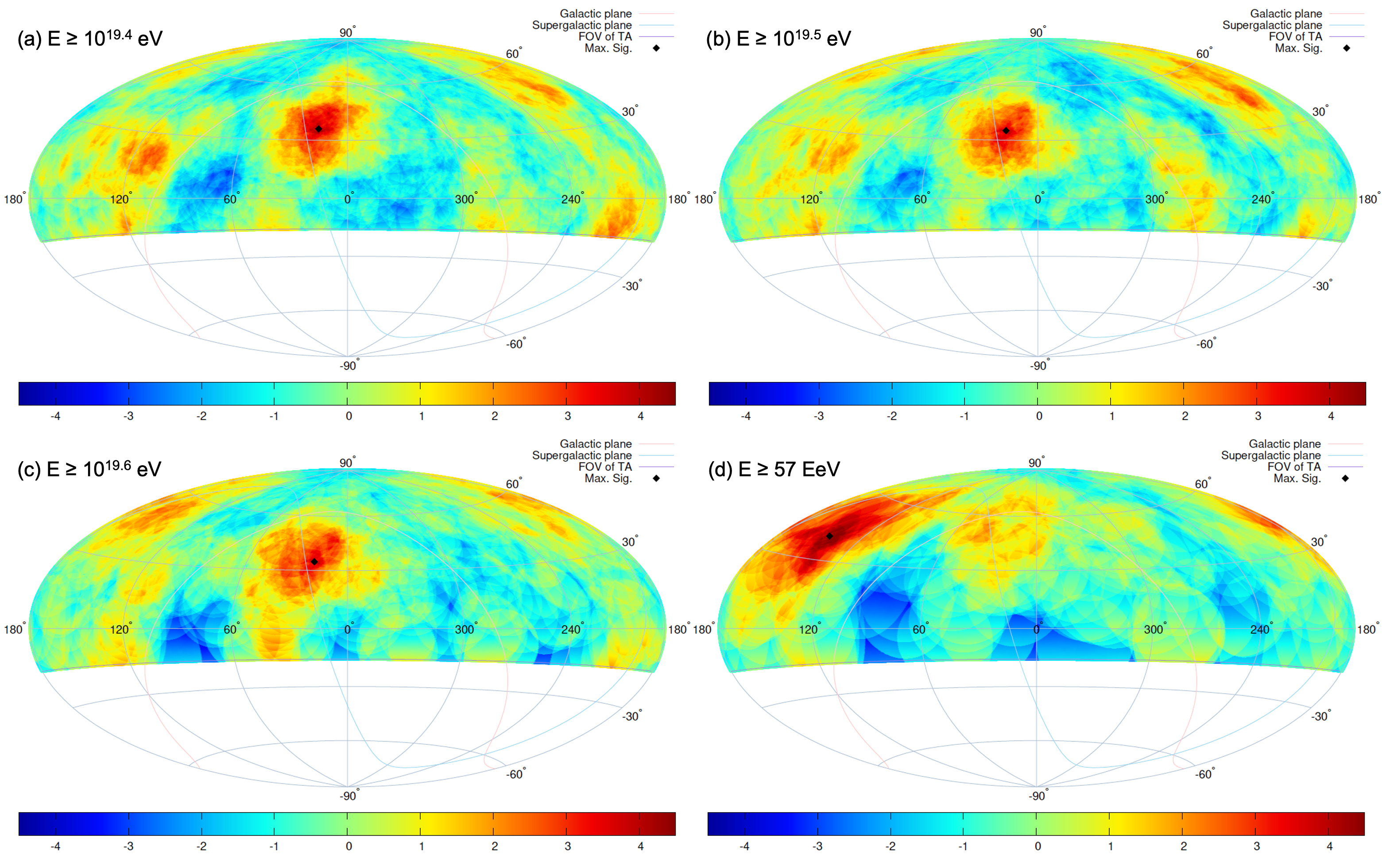}
    \caption{Sky maps in equatorial coordinates using Hammer projections. The color 
    scheme indicates the Li-Ma significance and shows the excess (red) or deficit 
    (blue) of events compared to isotropy at each grid point. The positions of 
    maximum excesses are marked with the black diamonds. An intermediate angular 
    scale of $20\degree$-radius circles was used for oversampling analysis for 
    different energy thresholds. The energy cut for each map is (a) $E \geq 10^{19.4} ~\eV$, 
    (b) $E \geq 10^{19.5} ~\eV$, (c) $E \geq 10^{19.6} ~\eV$, and (d) $E \geq 57 ~\EeV$. 
    For three lower energies, (a) to (c), a new excess of events is consistently observed 
    in the same direction. When the energy cut is raised to $E \geq 57 ~\EeV$ (d), the 
    maximum significance moves to the previously observed hotspot, however, a smaller 
    excess remains. Note that the right ascension of $0\degree$ is at the center of the sky map.}
    \label{fig:four_skymaps}
\end{figure*}

Figure \ref{fig:four_skymaps} shows the excesses of events with 
$E \geq 10^{19.4} ~\eV, 10^{19.5} ~\eV, 10^{19.6} ~\eV$, and $57 ~\EeV$, respectively. 
The color scheme represents the Li-Ma significance explained above. On each map, 
we indicated the position of maximum Li-Ma significance, excess with respect to isotropy, 
with a black diamond. For the three lower energies, the new excesses of events appear 
consistently in the direction to the left of the center in the sky maps in Figure 
\ref{fig:four_skymaps}. For the highest energy set, with $E \geq 57 ~\EeV$, the maximum 
significance of the sky map shifts to the region in the figure which is the hotspot 
reported in 2014 \citep{TA+hotspot:2014}. A smaller excess remains visible in the direction 
of the maximum at lower energies. 

\begin{figure*}[t]
    \centering
    \includegraphics[width=\textwidth]{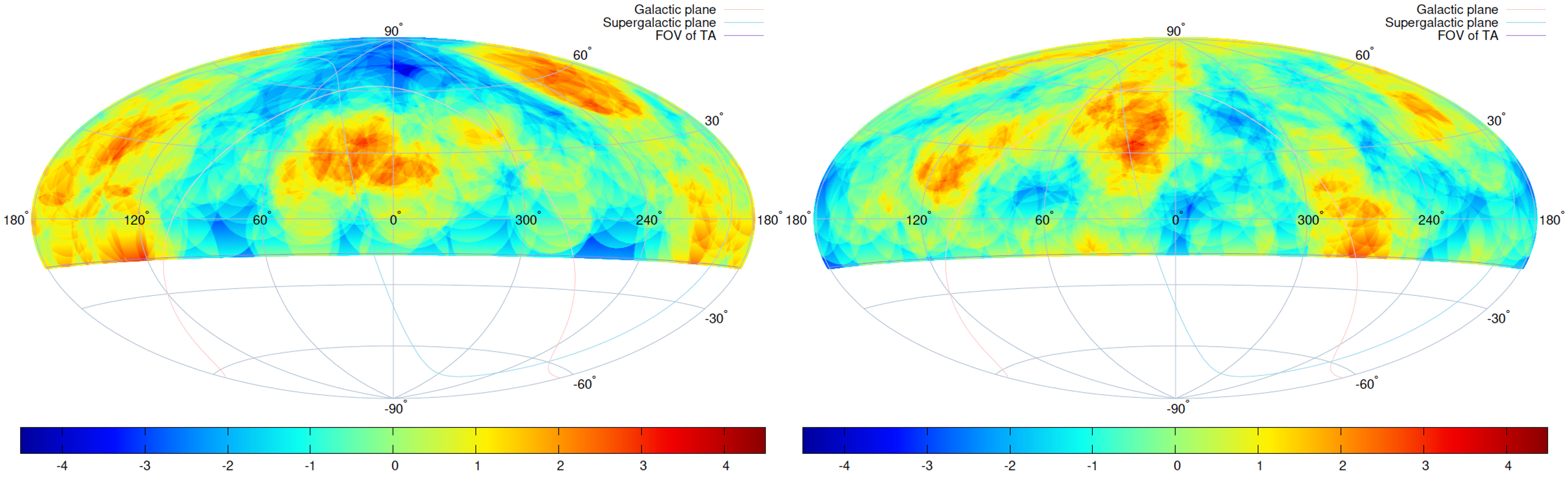}
    \caption{Sky maps in equatorial coordinates. The Li-Ma significance analysis maps 
    for the data set with $E \geq 10^{19.5} ~\eV$. The results for the first 5-years 
    of data are shown in the left and for the last 6-years of data on the right. 
    The same color scheme as Figure \ref{fig:four_skymaps} is used. Both maps have 
    a similar, $\sim$$3\sigma$, excess in new region which indicates that it may be 
    a steady state.}
    \label{fig:first_last_skymap}
\end{figure*}

Next, we investigated the time variation of the excess by dividing the data into two time 
periods. As an example, the sky map is shown in Figure \ref{fig:first_last_skymap} for 
the data set with $E \geq 10^{19.5} ~\eV$. The map is shown for the first 5-years of data 
adjacent to the map for the last 6-years of data. There is no apparent difference in the excess 
between the maps of the two data sets. Both maps have similar local significances, around $3\sigma$, 
toward the new excess region. The distributions of the other two energy cuts demonstrate similar 
local significances. This is indicative of a steady state excess in this region.

The new excess of events appears in the region of the Perseus-Pisces supercluster (PPSC), 
which is one of the notable structures within TA’s field of view. This is a possible 
source for the observed excess. The PPSC is the closest supercluster other than 
the local supercluster which we reside in. It is known that the PPSC has a gigantic 
filamentary structure, stretching for over $300 ~\Mpc/h$ \citep{Batuski:1985}. According 
to Haynes and Giovanelli \citep{Haynes+Giovanelli:1986}, the major portion of the 
foreground between the Earth and the PPSC as well as the space beyond the PPSC in the 
same direction, are nearly empty. Courtois \etal  provide density contour maps after 
making corrections for catalog incompleteness \citep{Courtois:2013}. Figure 8 in that paper 
shows the PPSC appears as an elongated structure of galaxies. No prominent structures 
between the Earth and the PPSC, or beyond the PPSC, are seen in the direction of the 
new excess of events in that figure. If the new excess of events in arrival direction 
distribution has an astrophysical origin, the PPSC could be responsible for it.

\begin{figure*}[t]
    \centering
    \includegraphics[width=\textwidth]{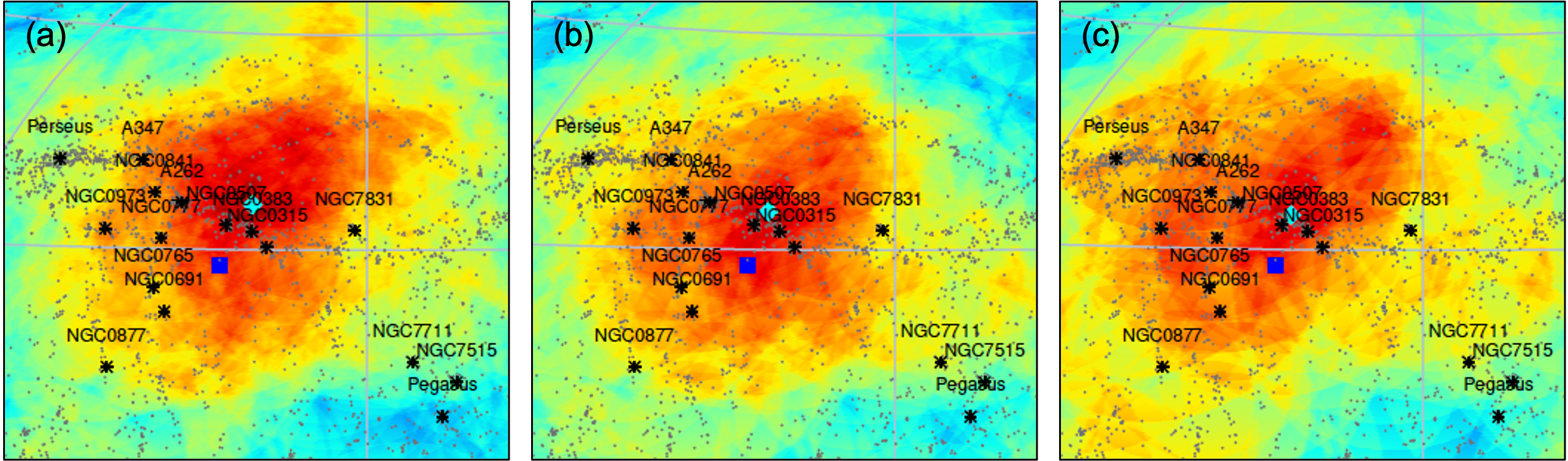}
    \caption{Expanded sky maps showing the new excesses of events overlaid with the major 
    clusters and groups of galaxies of the Perseus-Pisces supercluster (PPSC). The color 
    scheme is the same as that in Figure \ref{fig:four_skymaps}. The map is shown for three 
    energy thresholds with minimum energy increasing from left to right: (a) $E \geq 10^{19.4} ~\eV$, 
    (b) $E \geq 10^{19.5} ~\eV$, and (c) $E \geq 10^{19.6} ~\eV$. The representative elements 
    of the PPSC from reference \citep{PPSC} are indicated on the maps with black asterisks. 
    Galaxies from the 2MASS Redshift Survey catalog \citep{2MASS:2012}, with distances between 
    $35 ~\Mpc$ and $100 ~\Mpc$, are indicated with gray dots. These distances are similar to 
    those of the PPSC representative elements. The positions of maximum excesses are marked 
    with the cyan diamonds and the center of the PPSC is marked with the blue squares. It is 
    seen that the excess is coincident with the overall distribution of the PPSC. The angular 
    separations between the positions of the maximum excesses and the center of the PPSC are 
    less than $\sim$$10\degree$.}
    \label{fig:expanded}
\end{figure*}

Figure \ref{fig:expanded} shows the new excess in a series of expanded sky maps with 
the PPSC overlaid upon it. The data has been plotted for the three energy cuts using 
the Li-Ma significances. To show the elements of the PPSC on the sky, we adopt the 
list that includes the major clusters of galaxies and the major groups of galaxies that 
comprise the PPSC in reference \citep{PPSC}. These are marked with asterisks in the 
three maps. The excess is coincident with the overall distribution of the PPSC.

It is suggestive that the excess in the data falls on top of the PPSC. To quantify how often 
this happens by chance, we generate many isotropic MC event sets thrown according to the 
acceptance of the TA SD. Each MC set contains the same number of events as the data. We count 
as successes the number of MC sets where the maximum Li-Ma significance is at least as 
significant in the MC set as in the data and which occurs at least as close to the PPSC as the data.

We begin by defining the directional center of the PPSC using the mean values of the right 
ascensions and declinations of its representative elements. It is calculated to be 
($20.9\degree, 27.9\degree$) in equatorial coordinates, which is indicated by the blue square 
in Figure \ref{fig:expanded}. In addition, the center position of the excess, where the maximum 
Li-Ma significance is estimated, is represented by the cyan diamond in the figure. Next, we 
calculate the angular distances between the center of the PPSC and the positions of the maximum 
excesses— ($17.4\degree, 36.0\degree$), ($19.0\degree, 35.1\degree$), and ($19.7\degree, 34.6\degree$) 
for each energy threshold. Their angular separation is $8.6\degree$ for $10^{19.4} ~\eV$, $7.4\degree$ 
for $10^{19.5} ~\eV$, and $6.8\degree$ for $10^{19.6} ~\eV$.

The steps for performing the MC simulations are as follows. At each energy threshold, 
$10^{19.4} ~\eV$, $10^{19.5} ~\eV$, and $10^{19.6} ~\eV$, we throw $5 \times 10^5$ sets 
of MC trials with the same statistics as the data and perform 
a Li-Ma analysis of each trial, which gives us the maximum Li-Ma significance and its position. 
Then, we calculate the angle, $\theta_{\rm mc}$, between the position that has the MC’s maximum 
Li-Ma significance and the center of the PPSC. We count as successes those within angle $\theta_{\rm obs}$ 
of the PPSC with an equal or greater significance than the data with the PPSC: 
($S_{\rm mc} \geq S_{\rm obs}$) and ($\theta_{\rm mc} \leq \theta_{\rm obs}$). By requiring the MC set 
to meet or exceed the conditions of the data, we estimate the probability of having an equal or 
greater excess coincident with the PPSC by chance. The chance probabilities of having an excess this 
significant overlapping the PPSC are estimated to be $3.6\sigma$, $3.6\sigma$, and $3.4\sigma$ for 
$E \ge 10^{19.4} ~\eV$, $E \ge 10^{19.5} ~\eV$, and $E \ge 10^{19.6} ~\eV$, respectively.

\begin{figure*}[t]
    \centering
    \includegraphics[scale=0.6]{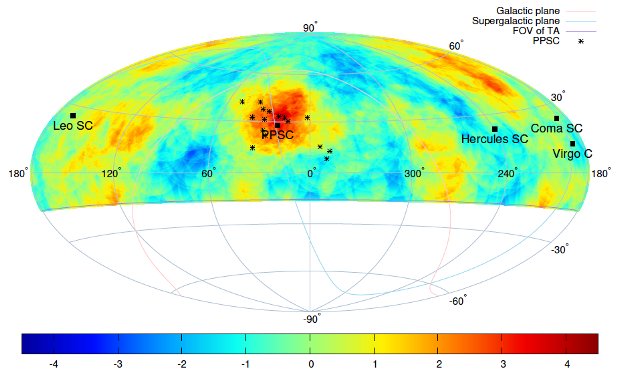}
    \caption{Sky map in equatorial coordinates. The Li-Ma significance analysis map for 
    the data set with $E \ge 10^{19.5} ~\eV$. The color scheme is the same as that in 
    Figure \ref{fig:four_skymaps}. The representative clusters and groups of galaxies 
    of the Perseus-Pisces supercluster (PPSC) are indicated by the black asterisks. The 
    nearby major structures within the Telescope Array field of view—Virgo cluster ($17 ~\Mpc$), 
    PPSC ($70 ~\Mpc$), Coma supercluster ($90 ~\Mpc$), Leo supercluster ($135 ~\Mpc$), and 
    Hercules supercluster ($135 ~\Mpc$)—are indicated by black squares. The data do not show an 
    excess at any of the locations of other nearby major structures other than the PPSC. 
    None of them have Li-Ma significances greater than $1\sigma$.}
    \label{fig:centers}
\end{figure*}

We investigate the data further by taking into account the major objects in the local large-scale 
structure of the universe similar to the PPSC within the TA's field of view: the Virgo cluster 
($17 ~\Mpc$), Coma supercluster ($90 ~\Mpc$), Leo supercluster ($135 ~\Mpc$), and Hercules 
supercluster ($135 ~\Mpc$). Their center positions are defined in the same manner as defining 
the PPSC center, which are marked with the black squares in Figure \ref{fig:centers}.
As an example, the Li-Ma significance map for the data set with $E \geq 10^{19.5} ~\eV$ is shown. 
The data do not show an excess at any of the locations of other major objects. All Li-Ma 
significances are less than $1\sigma$. 

However, to see if it is likely that an excess at one of their locations could occur by 
chance, we repeated the MC calculation described in the previous paragraph, replacing 
the PPSC with all the major structures listed here including the PPSC. The chance probability 
of having an excess of equal or greater significance than that of the data on top of any of the 
five major structures are estimated to be $3.1\sigma$, $3.0\sigma$, and $2.9\sigma$ for 
$E \geq 10^{19.4} ~\eV$, $E \geq 10^{19.5} ~\eV$, and $E \geq 10^{19.6} ~\eV$, respectively. 
These significances are sufficiently similar to that of the PPSC alone (about $3.5\sigma$) 
that we conclude that random coincidences with the major objects of the local large-scale 
structure are unlikely.

We repeated the calculation using NASA/IPAC Extragalactic Database \citep{NED} centers 
of the major structures, and the significances were the same. The results are summarized 
in Table \ref{tab:sum_chance_prob}. Random coincidences between the data and the PPSC occur 
at the $\sim$$3.5\sigma$ level. This result indicates that it is likely there has a cosmic 
ray source in the Perseus-Pisces supercluster.

\begin{table*}[!ht]
    \centering
    \caption{Summary of the Monte-Carlo studies that estimate the chance probability 
    of having an excess}    
    \begin{tabular}{ | c | c | c | c | c | }
        \hline
        \textbf{Energy (eV)} & \textbf{Events} & \textbf{Criteria} & \textbf{Perseus-Pisces}  & \textbf{Any of the Five}  \\
        & & & \textbf{supercluster} & \textbf{Major structures} \\
        \hline
        $E \geq 10^{19.4}$ & 864 & $4.4\sigma {\&} 8.6\degree$ & $3.6\sigma$ & $3.1\sigma$ \\ 
        \hline
        $E \geq 10^{19.5}$ & 558 & $4.2\sigma {\&} 7.4\degree$ & $3.6\sigma$ & $3.0\sigma$ \\
        \hline
        $E \geq 10^{19.6}$ & 335 & $4.0\sigma {\&} 6.8\degree$ & $3.4\sigma$ & $2.9\sigma$ \\     
        \hline
    \end{tabular}
    \label{tab:sum_chance_prob}
\end{table*}

\section{Summary} \label{sec:summary}
The TA Collaboration has observed a new excess of events in the arrival direction 
distribution. We found the excess over the isotropic background to have local 
significances of $4.4\sigma$, $4.2\sigma$, and $4.0\sigma$ for events of energy 
$E \geq 10^{19.4} ~\eV$, $E \geq 10^{19.5} ~\eV$, and $E \geq 10^{19.6} ~\eV$, respectively, 
by using the Li-Ma method and a $20\degree$-radius circle oversampling analysis. This 
excess overlaps with the Perseus-Pisces supercluster which is a nearby element of the 
local large-scale structure of the universe and is the closest supercluster to us 
(other than the Virgo supercluster within which we reside). 

When looking at the data overlaid with the PPSC, the excess is coincident with the 
overall distribution of the clusters and groups of galaxies within the Perseus-Pisces 
supercluster. To determine the probability that the data’s $4.0\sigma - 4.4\sigma$ 
significances could occur by chance close by the Perseus-Pisces supercluster, we 
generated isotropic Monte Carlo event sets with the same statistics as the data, 
thrown according to the acceptance of the TA surface detector. The chance probability 
of the excess of events occurring coincident with the Perseus-Pisces supercluster 
has $3.5\sigma$ significance.

We investigated whether there is another excess close to the locations of any of the 
nearby major structures similar to the Perseus-Pisces supercluster. None of them have 
Li-Ma significances larger than $1\sigma$. We repeated this process, testing the 
Monte Carlo event sets against the five nearby major astronomical structures. The 
significance of a random coincidence with any of them is estimated to be $\sim$$3\sigma$.

The excess of events observed in the direction of the Perseus-Pisces supercluster 
indicates that a cosmic ray source likely exists in that supercluster. The supercluster 
contains many interesting astronomical objects, including active galaxies, starburst 
galaxies, and large-scale shocks, that may be UHECR sources. It is important to study 
these astronomical objects in the supercluster further, and to increase the statistical 
power of Northern Hemisphere cosmic ray studies. 

The recent TA$\times$4 project, which is the extension of the TA SD aperture by a factor
of 4 and includes two new fluorescence detector stations which overlook the TA$\times$4 
SD array, is designed to study the highest energy cosmic rays \citep{TAx4:2021}. As of 2021, 
more than half of the TA$\times$4 SDs have been deployed and are operating successfully. 
Completing the construction of TA$\times$4 and continuing to run TA will likely be an 
essential key to solving the problem of the origin of ultrahigh energy cosmic rays.

A table of events with energies above $10^{19.4} ~\eV$ is available in a machine-readable 
form in the online journal. It includes date and time, zenith angle, right ascension, 
and declination.

%%\begin{acknowledgments}
%%\end{acknowledgments}
\section*{Acknowledgments}
The Telescope Array experiment is supported by the Japan Society for
the Promotion of Science(JSPS) through
Grants-in-Aid
for Priority Area
%"Highest Energy Cosmic Rays"
431,
for Specially Promoted Research
%``Extreme Phenomena in the Universe Explored by Highest Energy Cosmic Rays''
%Grant Number
JP21000002,
%Grant-in-Aid
for Scientific  Research (S)
%"Quest for the unified picture of the explosion mechanism of supernovae and the central engine of gamma-ray bursts"
%Grant Number
JP19104006,
%Grant-in-Aid
for Specially Promoted Research
%"Extended Telescope Array Experiment - Nearby Extreme Universe Elucidated by Highest-energy Cosmic Rays"
%Grant Number
JP15H05693,
%Grant-in-Aid
for Scientific  Research (S)
%"Study of the ultra high energy cosmic ray source evolution by detailed measurement of cosmic rays in the wide energy range"
%Grant Number
JP15H05741, for Science Research (A) JP18H03705,
%Grant-in-Aid
for Young Scientists (A)
%"hoge hoge"
%Grant Number
JPH26707011,
and for Fostering Joint International Research (B)
%"Search for Ultra-High Energy Cosmic Ray origin using the extended Telescope Array experiment"
%Grant Number
JP19KK0074,
by the joint research program of the Institute for Cosmic Ray Research (ICRR), The University of Tokyo;
by the Pioneering Program of RIKEN for the Evolution of Matter in the Universe (r-EMU);
by the U.S. National Science
Foundation awards PHY-1404495, PHY-1404502, PHY-1607727, PHY-1712517, PHY-1806797, PHY-2012934, and PHY-2112904;
by the National Research Foundation of Korea
% \linebreak
(2017K1A4A3015188, 2020R1A2C1008230, \& 2020R1A2C2102800) ;
%\linebreak
by the Ministry of Science and Higher Education of the Russian Federation under the contract 075-15-2020-778, IISN project No. 4.4501.18, and Belgian Science Policy under IUAP VII/37 (ULB). This work was partially supported by the grants ofThe joint research program of the Institute for Space-Earth Environmental Research, Nagoya University and Inter-University Research Program of the Institute for Cosmic Ray Research of University of Tokyo. The foundations of Dr. Ezekiel R. and Edna Wattis Dumke, Willard L. Eccles, and George S. and Dolores Dor\'e Eccles all helped with generous donations. The State of Utah supported the project through its Economic Development Board, and the University of Utah through the Office of the Vice President for Research. The experimental site became available through the cooperation of the Utah School and Institutional Trust Lands Administration (SITLA), U.S. Bureau of Land Management (BLM), and the U.S. Air Force. We appreciate the assistance of the State of Utah and Fillmore offices of the BLM in crafting the Plan of Development for the site.  Patrick A.~Shea assisted the collaboration with valuable advice and supported the collaboration’s efforts. The people and the officials of Millard County, Utah have been a source of steadfast and warm support for our work which we greatly appreciate. We are indebted to the Millard County Road Department for their efforts to maintain and clear the roads which get us to our sites. We gratefully acknowledge the contribution from the technical staffs of our home institutions. An allocation of computer time from the Center for High Performance Computing at the University of Utah is gratefully acknowledged.



%%\appendix

%%\section{List of Events}
%%List of events


\bibliography{main}{}
\bibliographystyle{aasjournal}



\end{document}